\documentclass[twocolumn,showpacs,aps,pra,amsmath,amssymb,superscriptaddress]{revtex4-1}
\usepackage{bm,color,bbm}
\usepackage{hyperref, mathtools,graphicx,soul}

\usepackage{subfigure}

\newcommand{\beq}{\begin{equation}}
\newcommand{\eeq}{\end{equation}}
\newcommand{\bqa}{\begin{eqnarray}}
\newcommand{\eqa}{\end{eqnarray}}
\newcommand{\nn}{\nonumber}

\newcommand{\rt}[1]{\sqrt{#1}\,}
\newcommand{\erf}[1]{Eq.~(\ref{#1})}

\newcommand{\smallfrac}[2]{\mbox{$\frac{#1}{#2}$}}
\newcommand{\bra}[1]{ \langle{#1} |}
\newcommand{\ket}[1]{ |{#1} \rangle}

\newcommand{\half}{\smallfrac{1}{2}}

\newcommand{\sq}[1]{\left[ {#1} \right]}

\newcommand{\tr}[1]{{\rm Tr}\sq{ {#1} }}

\newcommand{\xfrac}[2]{{#1}/{#2}}

\newcommand{\mf}{\mathbf}

\newcommand{\blk}{\color{black}}

\definecolor{maroon}{rgb}{0.7,0,0}

\definecolor{ngreen}{rgb}{0.3,0.7,0.3}

\definecolor{golden}{rgb}{0.8,0.6,0.1}

\definecolor{npurple}{rgb}{0.3,0,0.6}

\begin{document}
	\title{Volume monogamy of quantum steering ellipsoids for multi-qubit systems}
   \author{Shuming Cheng}
   \affiliation{Centre for Quantum Computation and Communication Technology (Australian Research Council), Centre for Quantum Dynamics, Griffith University, Brisbane, QLD 4111, Australia}
    \affiliation{Key Laboratory of Systems and Control, Academy of Mathematics and Systems Science, Chinese Academy of Sciences, Beijing 100190, P. R. China}

\author{Antony Milne}
\affiliation{Controlled Quantum Dynamics Theory, Department of Physics, Imperial College London, London SW7 2AZ, UK}

\author{Michael J. W. Hall}
\affiliation{Centre for Quantum Computation and Communication Technology (Australian Research Council), Centre for Quantum Dynamics, Griffith University, Brisbane, QLD 4111, Australia}

\author{Howard  M. Wiseman}
\affiliation{Centre for Quantum Computation and Communication Technology (Australian Research Council), Centre for Quantum Dynamics, Griffith University, Brisbane, QLD 4111, Australia}

	\begin{abstract}
The quantum steering ellipsoid can be used to visualise two-qubit states, and thus provides a generalisation of the Bloch picture for the single qubit. Recently, a monogamy relation for the volumes of steering ellipsoids has been derived for pure 3-qubit states and shown to be stronger than the celebrated Coffman-Kundu-Wootters (CKW) inequality. We first demonstrate the close connection between this volume monogamy relation and the classification of  pure 3-qubit states under stochastic local operations and classical communication (SLOCC). We then show that this monogamy relation does not hold for general mixed 3-qubit states and derive a weaker monogamy relation that does hold for such states.  We also prove a volume monogamy relation for pure 4-qubit states (further conjectured to hold for the mixed case), and generalize our 3-qubit inequality to $n$ qubits. 
Finally, we study the effect of  noise on the quantum steering ellipsoid and find that the volume of any two-qubit state is non-increasing when the state is exposed to arbitrary local noise. This  implies that any volume monogamy relation for a given class of multiqubit states remains valid under the addition of local noise. We investigate this quantitatively for the experimentally relevant example of isotropic noise. 

	\end{abstract}
	\maketitle
	\section{Introduction} \label{introduction}
	
	 Qubits play a fundamental role in quantum information processing tasks~\cite{NC00} and quantum measurement and control~\cite{WM09}. The Bloch vector faithfully captures all \blk features of a single qubit state. \blk However, finding an analogous geometric representation of multi-qubit states is harder.   An elegant solution to this problem has recently been given for the two-qubit case~\cite{JPJR14}. \blk  In particular,  for a two-qubit state shared between two parties, Alice and Bob say, \blk the  set of all possible \blk Bloch vectors that Alice \blk can steer Bob's \blk qubit to,  via \blk all possible local \blk measurements on her \blk qubit, forms an \blk ellipsoid, called the quantum steering ellipsoid. Together with Alice's and Bob's local Bloch vectors, \blk the quantum steering ellipsoid provides a geometric representation  of the shared \blk 2-qubit state~\cite{JPJR14}.  
		
    The set of two-qubit states has a rich structure, which is mirrored, and in some cases added to, by a corresponding zoo of quantum steering ellipsoids~\cite{JPJR14}.  For example, properties of steering ellipsoids reflect various features of  quantum correlations, such as discord~\cite{SJSD11, SSJYD12, JPJR14}, entanglement~\cite{JPJR14,  MJJWR14,MJJR14} and Einstein-Podolsky-Rosen (EPR)  \blk steering~\cite{SMMMH15, Quan16, NV16F, NV16S}.  Moreover, it appears that \blk steering ellipsoids may be useful \blk for characterising correlation properties of multiqubit states, such as monogamy, in new ways.  In particular, it  is well known that quantum correlations cannot be freely shared between members of multipartite systems, resulting in monogamy relations for, e.g., concurrence~\cite{CKW00, OV06},  Bell nonlocality~\cite{SG01,TV06, T09, KPRLK11}, and EPR-steering inequalities~\cite{R13}.  In this vein,  Milne {\it et al.}~ have recently obtained a strong monogamy relation for the volumes of the steering ellipsoids generated by  pure 3-qubit states, termed  volume monogamy~\cite{MJJWR14}, 
    which is strictly stronger than the well known Coffman-Kundu-Wootters (CKW) monogamy relation for concurrence~\cite{CKW00} in the pure regime.
  
 The volume monogamy relation for pure 3-qubit states  immediately suggests a number of questions: Does it discriminate between different types of entanglement? Is it valid for the mixed case? Are there similar  relations for multiqubit states? And what happens  when noise, induced by inevitable interaction with the environment and imperfections of local measurements, is present?  We  will  provide some answers to these questions in this paper, following \blk a brief review of the quantum steering ellipsoid and its properties in Sec.~\ref{II}.  
 
In Sec.~III we obtain and discuss volume monogamy relations in general scenarios ranging  from pure 3-qubit states to  general multiqubit states. First, in Sec.~III~A we discuss the known volume monogamy relation for pure 3-qubit states~\cite{MJJWR14}; give an alternative proof of this relation that demonstrates a link with the quantum marginal problem~\cite{K04}; review the derivation of the CKW monogamy relation from volume monogamy; and establish a close connection between properties of volume monogamy and the classification of pure 3-qubit states under stochastic local operations and classical communication (SLOCC). In Sec.~III~B  we show that this volume monogamy relation is  violated by some mixed states, and derive a weaker volume monogamy relation that is valid for all 3-qubit states,  pure or mixed.  We obtain a monogamy relation of the same form for pure 4-qubit states in Sec.~III~C,  which we conjecture is also applicable to mixed states, and give a nontrivial volume monogamy relation for general multiqubit states in Sec.~III~D.

  In Sec.~IV we investigate the effects of noise on steering ellipsoid volumes and on volume monogamy relations.  In  Sec.~IV~A  we show that  the volume of the steering ellipsoid decreases monotonically under arbitrary local operations, including under local noise channels in particular. This significantly generalises a recent result~\cite{HF15}, which is restricted to particular classes of states and types of noise. Moreover,  it implies that any volume monogamy relation for a given set of multiqubit states remains valid under the addition of local noise. This includes, in particular, the strong volume monogamy relation for pure 3-qubit states in Ref.~\cite{MJJWR14}. Finally, in Sec.~IV~B we investigate the experimentally relevant case of  isotropic noise acting on a family of 3-qubit W-class states. 

We conclude with some remarks and open questions in Sec.~\ref{V}.

\section{Quantum steering ellipsoids for  two-qubit states}\label{II}

	Any two-qubit state $\rho_{AB}$ can be written in the form 
	\begin{align}
		\rho_{AB} =  \frac{1}{4} \bigg( & \mathbbm{1}_A \otimes \mathbbm{1}_B+{\mf{a}} \cdot \boldsymbol{\sigma} \otimes \mathbbm{1}_B+\mathbbm{1}_A \otimes {\mf{b}} \cdot \boldsymbol{\sigma}     \nn \\  
						        &  +\sum^3_{j,k=1}T_{jk}\sigma_j\otimes\sigma_k \bigg), 
		\label{state}
	\end{align}
where $\boldsymbol{\sigma}\equiv (\sigma_1, \sigma_2, \sigma_3)$ denotes the vector of Pauli spin operators and $\mathbbm{1}_A, \mathbbm{1}_B$ are identity operators. Here ${\mf{a}}$ and ${\mf{b}}$ are the Bloch vectors of Alice's and Bob's reduced states and $T$ is the spin correlation matrix,  i.e.,
	\begin{align}
	 a_j&:=\tr{\rho_{AB}\sigma_j\otimes \mathbbm{1}_B}, ~~~ b_k:=\tr{\rho_{AB}\mathbbm{1}_A\otimes \sigma_k}, \nn \\
	 T_{jk}&:=\tr{\rho_{AB}\sigma_j\otimes \sigma_k}, \qquad (j, k=1, 2, 3). \label{notation}
	\end{align}

	Different choices of Alice's local measurements result in different {\it steered}  states for Bob.  Specifically, each local measurement outcome for Alice corresponds to some element $ E\geq 0$ of a positive operator-valued measure (POVM) describing her measurement. Thus, $E=e_0\left(\mathbbm{1}_A+\mf{e}\cdot\boldsymbol{\sigma} \right)$, with $e_0\geq 0$ and $|\mf{e}|\leq 1$. This outcome is obtained with 
probability  
	\beq
	p^E=\tr{\rho_{AB}\, E\otimes \mathbbm{1}_B}=e_0\left(1+{\mf{a}}\cdot \mf{e}\right), \nn
	\eeq
	leading to the steered state 
	\beq \label{reducedrho}
	\rho^E_B=\frac{{\rm Tr}_A[\rho_{AB}\,E\otimes \mathbbm{1}_B]}{p^E}=\frac{1}{2}\left(\mathbbm{1}_B+\frac{({\mf{b}}+T^\top\mf{e})\cdot \boldsymbol{\sigma}}{1+{\mf{a}}\cdot\mf{e}}\right) 
	\eeq
	for Bob's qubit.
	
	Considering all possible local measurements by Alice, it follows that the corresponding set of Bob's steered states is represented by the set of Bloch vectors
	\beq \label{reduced}
	 \mathcal{E}_{B|A}=\left\{ \frac{{\mf{b}}+T^\top\mf{e}}{1+{\mf{a}}\cdot\mf{e}}:|\mf{e}|\leq 1 \right\}.
	 \eeq
	While not immediately apparent from Eq.~(\ref{reduced}), this set forms a (possibly degenerate) ellipsoid, and hence is called a quantum steering ellipsoid~\cite{JPJR14}. The subscript $B|A$ indicates the steering ellipsoid for Bob that is generated by Alice's local measurements.  Similarly, there is a steering ellipsoid for Alice generated by Bob's local measurements, denoted by $\mathcal{E}_{A|B}$.
	The  ellipsoid $\mathcal{E}_{B|A}$ is fully determined by its centre,
	\beq
	{\mf{c}}_{B|A}=\frac{{\mf{b}}-T^\top{\mf{a}}}{1-a^2}, \nn
	\eeq
	and its orientation matrix,
	\beq
	Q_{B|A}=\frac{1}{1-a^2}\left(T-{\mf{a}}{\mf{b}}^\top\right)^\top\left(I+\frac{{\mf{a}}{\mf{a}}^\top}{1-a^2}\right)\left(T-{\mf{a}}{\mf{b}}^\top\right), \nn
	\eeq
	where the eigenvalues and corresponding eigenvectors of $Q_{B|A}$ describe the squared lengths of the ellipsoid's semiaxes and their orientations~\cite{JPJR14}. Here and elsewhere we use $x$ to denote $|\mf{x}|$ for the vector $\mf{x}$. 
	
	The quantum steering ellipsoid $\mathcal{E}_{B|A}$, together with the reduced Bloch vectors ${\mf{a}}$ and ${\mf{b}}$,  provides a faithful representation of any two-qubit state up to  local unitary operations on Alice's qubit ~\cite{JPJR14}. Additionally, its various geometric properties encode useful information about the state.
	 For example, the state is separable if and only if its steering ellipsoid can be nested in a tetrahedron that is in turn nested in the Bloch sphere~\cite{JPJR14}.  
	 
	The size of the steering ellipsoid is quantified by its volume~\cite{JPJR14}, 
	 \beq 
	 V_{B|A}=\frac{4\pi}{3}\frac{|\det (T-{\mf{a}}{\mf{b}}^\top)|}{(1-a^2)^2}. \label{volume}
	 \eeq
	 It is obvious that the steering ellipsoid is constrained to lie inside the Bloch sphere, implying the volume is always no larger than $ V_{{\rm unit}}=\frac{4\pi}{3}$.  It is therefore convenient to work with the corresponding {\it normalised volume} defined by
	 \beq \label{vnorm}
	 v_{B|A} := \frac{V_{B|A}}{V_{\rm unit}} \leq 1.
	 \eeq \blk
	 The upper bound is achieved if and only if Alice and Bob share a pure entangled 2-qubit state~\cite{JPJR14},  and hence the steering ellipsoids of such states coincide with the Bloch ball.  In contrast, the normalised volume of any separable state is restricted by the nested tetrahedron condition to be no greater than $\xfrac{1}{27}$~\cite{JPJR14}. Thus, the steering ellipsoid volume is, at least to some extent, connected with the entanglement of the state. This paper will explore this connection further, via volume monogamy relations.

For later reference, we note here that the quantum steering ellipsoid $\mathcal{E}_{B|A}$, and hence its volume $V_{B|A}$, is invariant under the local filtering operation on Alice's qubit defined by \cite{MJJWR14}
\beq \label{filter}
\tilde \rho_{AB}:=\left[(2\rho_A)^{-1/2}\otimes \mathbbm{1}_B \right]\rho_{AB} \left[ (2\rho_A)^{-1/2}\otimes \mathbbm{1}_B \right] .
\eeq
The filtered state $\tilde\rho_{AB}$ is called the canonical form of $\rho_{AB}$, and has the useful properties $\tilde{\mathcal{E}}_{B|A}=\mathcal{E}_{B|A}$ and $\tilde{{\mf{a}}}=\mf{0}$.  Thus, for example, the normalised volume of $\mathcal{E}_{B|A}$ can be rewritten in the simple form
\beq \label{vcan}
v_{B|A} = \tilde v_{B|A} = |\det \tilde T_{AB}|,
\eeq
via Eqs.~(\ref{volume}) and~(\ref{vnorm}), where $\tilde T_{AB}$ denotes the spin correlation matrix for $\tilde\rho_{AB}$.
Note that the canonical form is well-defined whenever Alice does not have a pure qubit state, i.e., whenever $a\neq1$. Conversely, for $a=1$ the shared state factorises, and hence Bob's steering ellipsoid trivially reduces to the single point $\mathcal{E}_{B|A} =\{\mf{b}\}$, with $v_{B|A}=0$.

\section{Volume Monogamy}\label{III}

\subsection{Pure 3-qubit states}

\subsubsection{Monogamy, quantum marginals and concurrence}

	Consider now the scenario in which Alice, Bob, and Charlie share a pure 3-qubit state, $\rho_{ABC}=\ket{\psi_{ABC}}\bra{\psi_{ABC}}$, and let $\mathcal{E}_{B|A}$  and $\mathcal{E}_{C|A}$ denote the steering ellipsoids for Bob and Charlie, respectively, generated by local measurements on Alice's qubit. Milne {\it et al.} have derived the strong volume monogamy relation~\cite{MJJWR14,ASDHT15}
	\beq
	\rt{v_{B|A}}+\rt{v_{C|A}}\leq 1 \label{monogamy}
	\eeq
for the corresponding normalised volumes.

This relation immediately implies that Alice cannot steer both Bob and Charlie to a large set of states. For example, if Alice is able to steer Bob to the whole Bloch sphere (i.e., they share a pure entangled state), then Charlie's steering ellipsoid has zero volume (and indeed reduces to a single point).  The volume monogamy relation~(\ref{monogamy}) is depicted in Fig.~1. It is nontrivially saturated if and only if $\ket{\psi_{ABC}}$ is a W-class state~\cite{MJJWR14,ASDHT15} (see also Sec.~III~A.2 below). 

We note here an alternative proof of Eq.~(\ref{monogamy}) to that given in Ref.~\cite{ASDHT15}, which does not require consideration of extremal ellipsoid volumes, and which provides an interesting connection between volume monogamy and the quantum marginal problem~\cite{K04,HSS03}.
For $a=1$ the proof is trivial: Alice's state is pure and hence the shared state factorises, implying that $v_{B|A}=v_{C|A}=0$. Otherwise, for $a\neq1$ we can apply a local filtering operation similarly to Eq.~(\ref{filter}), with $\mathbbm{1}_B$ replaced by $\mathbbm{1}_{BC}$ and $\rho_{AB}$ replaced by $\rho_{ABC}$, to obtain the corresponding canonical state $\tilde \rho_{ABC}$ with $\tilde{\mf{a}}=\mf{0}$. Taking partial traces, the normalised volumes of Bob's and Charlie's steering ellipsoids follow via Eq.~(\ref{vcan}) as 
		\begin{align} \label{vba}
		v_{B|A}&=\tilde v_{B|A}=|\det \tilde T_{AB}| = \tilde c^2 , \\
		\label{vca}
	    v_{C|A}&=\tilde v_{C|A}= |\det \tilde T_{AC}| = \tilde b^2,
	    \end{align}
respectively, where the final equalities may be proved for pure canonical states by direct calculation,  or via partial transposition properties as per Ref.~\cite{ASDHT15}.
Finally, we apply the polygon inequality~\cite{HSS03,K04} 
	   \beq
	   b+c \leq 1+a \label{polygon} ,
	   \eeq
for the Bloch vectors of any pure 3-qubit state, to the canonical state $\tilde \rho_{ABC}$, yielding
	    \begin{align*}
	    \rt{v_{B|A}}+\rt{v_{C|A}} & =\tilde c + \tilde b\leq 1 + \tilde a = 1
	    \end{align*}  
as required.

Remarkably, the volume monogamy relation~(\ref{monogamy}) also has a deep connection with entanglement monogamy. In particular, the concurrence of a bipartite state $\rho_{AB}$ has the upper bound~\cite{MJJWR14}
   \beq 
   C^2(\rho_{AB})\leq (1-a^2)\rt{v_{B|A}}, \label{concurrence}
   \eeq
which combined with Eq.~(\ref{monogamy}) immediately yields the celebrated CKW  inequality~\cite{CKW00},  
   \beq
   C^2(\rho_{AB})+C^2(\rho_{AC})\leq (1-a^2)=4\det \rho_A . 
   \eeq
Thus, volume monogamy is a mathematically stronger condition than the monogamy of concurrence. Note, however, the latter is valid for any 3-qubit state, including mixed states. This is not the case for Eq.~(\ref{monogamy}), as we will show in Sec. III B.

\begin{figure}\label{Fig1}
	\centering
	\includegraphics[width=0.40\textwidth]{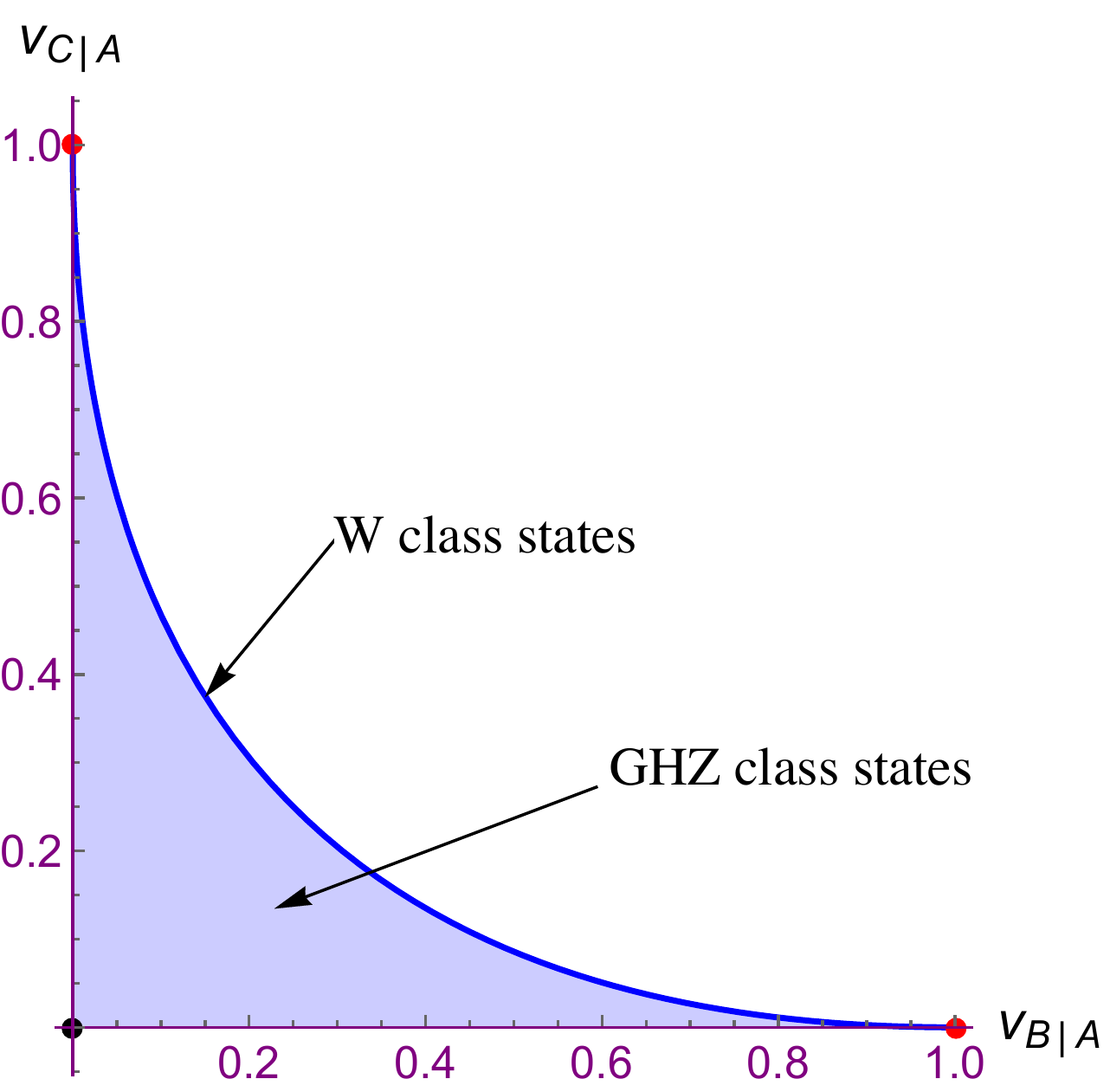}\\
	
	\caption{ The volume monogamy relation, \erf{monogamy}, for pure states. The $x$-axis and $y$-axis refer to the normalised volumes $v_{B|A}$ and $v_{C|A}$ respectively. Both fully factorisable states and bipartite entangled states $\ket{\psi_A}\ket{\psi_{BC}}$ are mapped to the origin.  Two red points $(1, 0)$ and $(0, 1)$ correspond to the other two classes of bipartite entangled states $\ket{\psi_{AB}}\ket{\psi_{C}}$ and $\ket{\psi_{AC}}\ket{\psi_{B}}$. The blue solid line is $\rt{v_{A|B}}+\rt{v_{C|B}}=1$ and represents the W-class states, except for the red points $(1, 0)$ and $(0, 1)$. The shaded  region, including the two axes, represents the GHZ-class states.} 
\end{figure}

\subsubsection{Connection to SLOCC classes and 3-tangle}

It has been shown in Ref.~\cite{DVC00} that any pure 3-qubit state can be classified into one of six entanglement classes,  with all members of any one class being interconvertible under SLOCC transformations, i.e., under local operations and classical communication with some nonzero probability. Here we show that these classes map onto corresponding regions of Fig.~1, thus relating these classes to properties of steering ellipsoid volumes.

We first consider the four classes having no tripartite entanglement.  The first of these comprises all fully factorisable states, i.e., those of the form $\ket{\psi_{ABC}}=\ket{\psi_A}\ket{\psi_B}\ket{\psi_C}$. All steering ellipsoids reduce to single points for this class, with zero volumes, and thus it is mapped to the origin in Fig~1. Next  are the three classes of bipartite entangled states, with the corresponding forms $|\psi_A\rangle|\psi_{BC}\rangle$, $|\psi_{AB}\rangle|\psi_C\rangle$ and $|\psi_{AC}\rangle|\psi_B\rangle$. For the first of these, Alice's qubit is uncorrelated with Bob's and Charlie's qubits, and hence the steering ellipsoids ${\cal E}_{B|A}$ and ${\cal E}_{C|A}$ again have zero volumes, corresponding to the origin in Fig.~1. For the other two bipartite classes, Alice can steer one of Bob and Charlie's qubits to the entire Bloch sphere, and the other to a single point. Thus, these two classes correspond to the two red dots in Fig.~1, and trivially saturate volume monogamy relation~(\ref{monogamy}).

The remaining two entanglement classes comprise genuine 3-party entangled states. They correspond to states which are convertible, under SLOCC transformations, to either  the W-state
		\beq
		\ket{\psi_{ABC}}=\frac{1}{\rt{3}}\left(\ket{100}+\ket{010}+\ket{001}\right),
		\eeq
or to the Greenberger-Horne-Zeilinger (GHZ) state
\beq
\ket{\psi_{ABC}}=\frac{1}{\rt{2}}\left(\ket{000}+\ket{111}\right),
\eeq 
and are called the W-class and the GHZ-class, respectively~\cite{DVC00}.  These two classes are distinguished by having 3-tangles $\tau(\rho_{ABC})=0$ and $\tau(\rho_{ABC})>0$, respectively~\cite{DVC00}, where for pure states ~\cite{CKW00}
\beq \label{tangle}
\tau(\rho_{ABC}):=4\det\rho_A-C^2(\rho_{AB})-C^2(\rho_{AC}). 
\eeq
As indicated in Fig.~1, and demonstrated below, W-class states are precisely those states which nontrivially saturate the volume monogamy relation , while the GHZ-class states correspond to the entire region below this saturating curve. 

It has been shown by Milne {\it et al.} that the volume monogamy relation is saturated if and only if the state is a `maximum volume' state, with the canonical form
\beq
|\tilde\psi_{ABC}\rangle = \left( |100\rangle + \cos\theta|010\rangle + \sin\theta|001\rangle\right)/\sqrt{2} 
\eeq
up to local unitary transformations, where $\theta\in[0,\pi/2]$~\cite{MJJWR14,ASDHT15}. It is straightforward to check that such states are bipartite entangled states for $\theta=0,\pi/2$, corresponding to the points $(0,1)$ and $(1,0)$ in Fig.~1, and are W-class states otherwise~\cite{DVC00}.  Hence, all states nontrivially saturating the volume monogamy relation are W-class states.  Conversely, noting that W-class states have zero 3-tangle and $a<1$~\cite{DVC00}, the inequality
\[ \tau(\rho_{ABC}) \geq (1-a^2) \left(1-\sqrt{v_{B|A}}-\sqrt{v_{C|A}} \right) \]
following from Eqs.~(\ref{concurrence}) and~(\ref{tangle}) immediately implies that every W-class state saturates the monogamy relation. 

It follows from the above that all GHZ-class states must correspond to points in the region below the saturating curve in Fig.~1.  We show that indeed every point in this region, including the axes, corresponds to such a state.  In particular, we give a family of canonical GHZ-class states for which the normalised volumes are mapped to every point $(x, y)$ in the shaded area in Fig.~1, including the axes. The explicit form of these states is defined via two real free parameters:
\begin{align}
\ket{\tilde \psi_{ABC}}=\frac{1}{\rt{2}}&(\sin \alpha \ket{100}+\sin \beta\ket{010} \nn \\ &+\cos\beta\ket{001}+\cos\alpha\ket{111}),
\end{align}
where $\alpha, \beta \in (0, \pi/2)$, which immediately maps to the coordinates $x=v_{B|A}=\frac{1}{4}(\cos 2\alpha+\cos2\beta)^2$ and $y=v_{C|A}=\frac{1}{4}(\cos2\alpha-\cos2\beta)^2$ via Eqs.~(\ref{vba}) and~(\ref{vca}).  These fill the shaded area because  any point $(x, y)$ therein corresponds to either:  $ 2\alpha=\arccos(\rt{x}+\rt{y})$ and  $2\beta=\arccos(\rt{x}-\rt{y})$; or $ 2\alpha=\arccos(\rt{x}-\rt{y})$ and  $2\beta=\arccos(\rt{x}+\rt{y})$.

 	\subsection{ Mixed 3-qubit states} \label{Mixed}

A natural question to consider is whether the volume monogamy relation~(\ref{monogamy}) is also valid for a general mixed 3-qubit state. Unfortunately, the answer is negative. A counterexample is given by
	\beq
	\rho_{ABC}=\frac{1}{2}\left(\ket{\chi_1} \bra{\chi_1 }+\ket{\chi_2} \bra{\chi_2 }\right), \label{counterexample}
	\eeq
with
	\begin{align*}
	\ket{\chi_1}&=\frac{1}{\rt{6}}\left(\ket{101}-2\ket{011}+\ket{110}\right), \\
	\ket{\chi_2}&=\frac{1}{\rt{6}}\left(\ket{010}-2 \ket{100}+\ket{001}\right).
	\end{align*}
The 2-qubit reduced states $\rho_{AB}$ and $\rho_{AC}$ are then identical Werner states, of the form
	\beq
    \frac{2}{3}\ket{\psi^-}\bra{\psi^-}+\frac{1}{3}\frac{\mathbbm{1}}{2}\otimes \frac{\mathbbm{1}}{2}, 
	\eeq
	where $\ket{\psi^- }$ denotes the singlet state $(\ket{01}-\ket{10})/\rt{2}$. It is easy to verify that the corresponding steering ellipsoids are spheres of radius $2/3$~\cite{JPJR14}. Hence, $\rt{v_{B|A}}+\rt{v_{C|A}}=2\,\rt{8/27}=1.0888> 1$, implying that the volume monogamy relation~(\ref{monogamy}) does not hold for all mixed 3-qubit states.

 In Sec.~IV we will show that Eq.~(\ref{monogamy}) does remain valid for the particular case of mixed states obtained via local operations on pure 3-qubit states. Here, however, we will derive  a volume monogamy relation that is valid for {\it all} 3-qubit states: 
   \beq
	 (v_{B|A})^{2/3} + (v_{C|A})^{2/3} \leq 1. \label{analytic}
	 \eeq	 
This monogamy relation is clearly weaker than Eq.~(\ref{monogamy}) for pure states. However, it is saturated in some cases---for example, when $\rho_{ABC} = \rho_{AB}\otimes \rho_C$ where $\rho_{AB}$ is a pure entangled state, in which case $v_{B|A}=1$ and $v_{C|A}=0$.

Our derivation of Eq.~(\ref{analytic}) is based on a relatively strong tradeoff relation
 for the pairwise spin correlations of any 3-qubit state $\rho_{ABC}$, 
 	\beq
 	\tr{T^\top_{AB}T_{AB}}+\tr{T^\top_{AC}T_{AC}}+\tr{T^\top_{BC}T_{BC}} \leq 3 \label{important}
 	\eeq
(with equality for pure states), obtained as follows. 
First, consider a pure state $\rho_{ABC}$. From the Schmidt decomposition, the purities of any bipartition of such a state satisfy
	 $\tr{\rho^2_{AB}}=\tr{\rho^2_C}$, 
	 $\tr{\rho^2_{AC}}=\tr{\rho^2_B}$,
	and $\tr{\rho^2_{BC}}=\tr{\rho^2_A}$,
which from the definitions in Eq.~(\ref{notation}) are equivalent to
	\begin{align*}
	\tr{T^\top_{AB}T_{AB}}+a^2+b^2&=1+2c^2, \\
	\tr{T^\top_{AC}T_{AC}}+a^2+c^2&=1+2b^2,\\
	\tr{T^\top_{BC}T_{BC}}+b^2+c^2&=1+2a^2.
	\end{align*}
Summing these three equations immediately leads to the identity
	\beq
	\tr{T^\top_{AB}T_{AB}}+\tr{T^\top_{AC}T_{AC}}+\tr{T^\top_{BC}T_{BC}}=3 \label{correlation}
	\eeq
for pure 3-qubit states.
Second, expressing a  mixed state $\rho_{ABC}$  as a convex combination of pure states, $\rho_{ABC}=\sum_m p_m\ket{\varphi_m}\bra{\varphi_m}$,
and letting $T^m_{AB}$ denote the spin correlation matrix of ${\rm Tr}_C{\ket{\varphi_m}\bra{\varphi_m}}$, we have
\begin{align}& \tr{T^\top_{AB}T_{AB}} \nn \\
&= \sum_{m,n}p_mp_n\tr{\left(T^m_{AB}\right)^\top T^n_{AB}} \nn\\
& \leq  \sum_{m,n}p_mp_n\left(\tr{\left(T^m_{AB}\right)^\top T^m_{AB}} \tr{\left(T^n_{AB}\right)^\top T^n_{AB}} \right)^{1/2} \nn \\ \nn
&\leq \frac{1}{2} \sum_{m,n}p_mp_n\left(\tr{\left(T^m_{AB}\right)^\top T^m_{AB}} +\tr{\left(T^n_{AB}\right)^\top T^n_{AB}} \right) \\
& = \sum_mp_m\tr{\left(T^m_{AB}\right)^\top T^m_{AB}}. 
\end{align}	 
Here the first inequality follows from the Schwarz inequality, and the second from the geometric mean being no greater than the arithmetic mean.  One has similar relations for $\tr{T^\top_{AC}T_{AC}}$ and $\tr{T^\top_{BC}T_{BC}}$. 
Summing these and using identity~(\ref{correlation}) for pure states then yields Eq.~(\ref{important}) as required.

We note that Eq.~(\ref{important}) can itself be considered as a monogamy relation, for the strengths of the pairwise spin correlations, and subsumes the known monogamy relation for pair-wise Bell nonlocality~\cite{QFL15}.

Similarly to the proof of Eq.~(\ref{monogamy}) in the previous section, we now consider the canonical state $\tilde \rho_{ABC}$ for $a\neq 1$ (since Eq.~(\ref{analytic}) is similarly trivially satisfied when $a=1$).
Equation~(\ref{vcan}) then yields 
\begin{align}
   (v_{B|A}&)^{2/3}+(v_{C|A})^{2/3} \nn \\
&=\left|\det \left[(\tilde T_{AB})^\top \tilde T_{AB}\right]\right|^{1/3}
 +\left|\det \left[(\tilde T_{AC})^\top \tilde T_{AC}\right]\right|^{1/3} \nn \\
&\leq \frac{1}{3}\tr{(\tilde T_{AB})^\top \tilde T_{AB}}+\frac{1}{3}\tr{(\tilde T_{AC})^\top \tilde T_{AC}} \nn \\ 
&\leq 1\label{0.66}
\end{align}
as desired.  Here the first inequality follows from the arithmetic-geometric mean inequality, applied to the eigenvalues of $(\tilde T_{AB})^\top \tilde T_{AB}$ and $(\tilde T_{AC})^\top \tilde T_{AC}$, and the second inequality from the tradeoff relation in Eq.~(\ref{important}).

In the next section we show that a monogamy relation of the same form (involving $2/3$ powers) also holds for pure 4-qubit states.
	
\subsection{4-qubit states}

We first remark that the strong monogamy relation~(\ref{monogamy}) for pure 3-qubit states cannot be generalised to a similar form for pure n-qubit states, for any $n\geq 4$. In particular, by purifying the counterexample in Eq.~(\ref{counterexample}), we can construct the pure 4-qubit state
	\beq 
	\ket{\psi}_{ABCD}=\frac{1}{\sqrt 2}\left(\ket{\chi_1}_{ABC}\ket{0}_D+ \ket{\chi_2}_{ABC}\ket{1}_D \right),
	\eeq 
	 implying, similarly to the counterexample, that $\rt{v_{B|A}}+\rt{v_{C|A}}+\rt{v_{D|A}}=2\rt{8/27}+\rt{v_{D|A}}> 1$.
	 It follows more generally, by considering an n-qubit pure state with factor $\ket{\psi}_{ABCD}$, that the form of Eq.~(\ref{monogamy}) does not generalise to any $n\geq 4$.
	
We will show here, however, that the volume monogamy relation
	\beq
(v_{B|A})^{2/3}+(v_{C|A})^{2/3} + (v_{D|A})^{2/3}\leq 1 \label{4qubit}
   \eeq	
is valid for any pure 4-qubit state $\rho_{ABCD}$, and give numerical evidence strongly supporting its validity for mixed 4-qubit states.\blk
	
To prove result (\ref{4qubit}), we adapt the techniques used in the proof of Eq.~(\ref{analytic}) in Sec.~\ref{Mixed}. First, using the equalities of purities of any bipartition of a pure state, we have 
	{\small \begin{align}
	a^2+b^2+\tr{T^\top_{AB}T_{AB}}&=c^2+d^2+\tr{T^\top_{CD}T_{CD}} , \label{AB} \\
	a^2+c^2+\tr{T^\top_{AC}T_{AC}}&=b^2+d^2+\tr{T^\top_{BD}T_{BD}} ,\label{AC} \\
	a^2+d^2+\tr{T^\top_{AD}T_{AD}}&=b^2+c^2+\tr{T^\top_{BC}T_{BC}}, \label{AD}
	\end{align}}
with respect to the bipartitions $(AB,CD)$, $(AC,BD)$ and $(AD,BC)$. We also have, for the bipartition $(A,BCD)$,
\begin{align}
b^2+c^2+d^2&+\tr{T^\top_{BC}T_{BC}+T^\top_{BD}T_{BD}+T^\top_{CD}T_{CD}} \nn \\
	                        &+L_{BCD}=3+4a^2,\label{BCD} 
\end{align}
where
\beq
L_{BCD}:=\sum_{l,m,n}\big(\tr{\mathbbm{1}\otimes\sigma_l\otimes\sigma_m\otimes\sigma_n\,\rho_{ABCD}}\big)^2
\eeq
 is a measure of the tripartite spin correlation strength between Bob, Charlie and Dianne.
Summing Eqs.~(\ref{AB})-(\ref{BCD}) yields
	\begin{align}
	   &\tr{(T^\top_{AB}T_{AB}}+\tr{T^\top_{AC}T_{AC}}+\tr{T^\top_{AD}T_{AD}} \nn \\
	 &= 3+a^2-L_{BCD} \leq 3+a^2.
	 \end{align}
Applying a local filtering operation similarly to Eq.~(\ref{filter}), with $\mathbbm{1}_B$ replaced by $\mathbbm{1}_{BCD}$ and $\rho_{AB}$ replaced by $\rho_{ABCD}$, to obtain the corresponding canonical state $\tilde\rho_{ABCD}$,  we have $\tilde{\mf{a}}=\mf{0}$ and hence that 
{\small
\beq
	\tr{(\tilde T_{AB})^\top \tilde T_{AB}}+\tr{(\tilde T_{AC})^\top \tilde T_{AC}}+\tr{(\tilde T_{AD})^\top \tilde T_{AD}}\leq  3.\label{sum}
\eeq }
Eq.~(\ref{4qubit}) then follows via the same arguments used in the derivation of Eq.~(\ref{0.66}).

	 Finally, we conjecture that  inequality ~(\ref{sum}) generalises to
		\beq
		\tr{T_{AB}T^\top_{AB}}+	\tr{T_{AC}T^\top_{AC}}+	\tr{T_{AD}T^\top_{AD}}\leq 3. \label{conjecture}
		\eeq		
		 for all pure 4-qubit states.  We have employed numerical simulations to generate $2\times10^5 $ random pure states and  find no violation of inequality (\ref{conjecture}). The validity of this conjecture would imply, using the same techniques as above, that monogamy relation~(\ref{4qubit}) in fact holds for {\it all} 4-qubit states. \blk

\subsection{Multiqubit states}

We now obtain a general volume monogamy relation for $n$-qubit states, pure or mixed, based on Eq.~(\ref{analytic}) for 3-qubit states. In particular, for an $n$-qubit state $\rho_{ABCD\dots}$ consider the normalised volumes $v_{B|A}, v_{C|A}, v_{D|A}, \dots$ of the steering ellipsoids generated by Alice's local measurements. The steered parties $B, C, D, \dots$ can be grouped into $\half(n-1)(n-2)$ distinct pairs, with each pair satisfying a volume monogamy relation as per Eq.~(\ref{analytic}).  Summing these relations over all such pairs and rearranging terms then yields the general monogamy relation
      \beq
	(v_{B|A})^{2/3}+(v_{C|A})^{2/3}+(v_{D|A})^{2/3}+\dots~ \leq \frac{n-1}{2}. \label{nqubit}
		\eeq
 This reduces to the 3-qubit relation for $n=3$, and in general places a nontrivial constraint on the degree to which Alice can steer the states of the other parties.  For example, noting that $v\leq v^{2/3}$ for $v\leq 1$, it follows that the average volume of the $n-1$ ellipsoids to which Alice can steer the other parties is bounded by
\beq
\bar v_{~ |A}:=\frac{v_{B|A}+v_{C|A}+v_{D|A}+\dots}{n-1}~ \leq \half. 
\eeq
	
For the 4-qubit case, the upper bound in Eq.~(\ref{nqubit}) is $3/2$. While this is  weaker than the upper bound of 1 in Eq.~(\ref{4qubit}) for pure 4-qubit states, it has the advantage of also being valid for the mixed case.

	\section{Volume monogamy and noise} \label{IV}
	
	\subsection{Local noise and steering ellipsoids}
	Taking into account the imperfections of any experiment, including in state preparation and measurement, the quantum state is inevitably exposed to all kinds of noise. Such noise processes can be modeled as a noisy channel acting on an ideal initial state.  We are interested in the problem of how  such channels affect the desired properties of the initial state, and in particular the steering ellipsoid. 
	
	Mathematically, a noisy channel acting on a bipartite state $\rho_{AB}$  is equivalent to a completely positive and trace preserving (CPTP) map, $\Phi$, mapping the initial state $\rho_{AB}$ to some $\rho^\prime_{AB}$. 
	 Here, we consider the case that the noise acts locally on each subsystem. Thus, $\Phi=\phi_A\otimes\phi_B$ where $\phi_A$ and $\phi_B$ are CPTP maps acting on  $A$ and $B$, respectively, and
	\beq
	 \rho^\prime_{AB}=\Phi(\rho_{AB})=(\phi_A\otimes \phi_B)(\rho_{AB}).
	 \eeq
	
The set of reduced states generated by Alice's local measurements on $\rho^\prime_{AB}$ follows from Eq.~(\ref{reducedrho}) of Sec.~II as 
	\begin{align}
\{\rho^{\prime\,E}_B\}&=\left\{ \frac{{\rm Tr}_A\left[ (\phi_A\otimes\phi_B)(\rho_{AB})\,E\otimes \mathbbm{1}_B\right]}{\tr{(\phi_A\otimes\phi_B)(\rho_{AB})\,E\otimes \mathbbm{1}_B}}\right\} \nn \\ 
	&=  \left\{ \frac{{\rm Tr}_A\left[ (I_A\otimes \phi_B)(\rho_{AB})\, \phi_A^\star(E)\otimes \mathbbm{1}_B \right]}{\tr{(I_A\otimes \phi_B)(\rho_{AB})\, \phi_A^\star(E)\otimes \mathbbm{1}_B}}\right\} , 
	  \nn
	\end{align}
where $E$ ranges over all positive operators, $I$ denotes the identity map, and the dual map $\phi^\star$ of any CP map $\phi$ is defined  by $\tr{\phi^\star(X)Y}:=\tr{X\phi(Y)}$.
Noting $\phi_B$ is trace preserving and that $\phi^\star_A$ maps positive operators to positive operators then yields  
	\begin{align}
	\{\rho^{\prime\,E}_B\}
		&=\left\{ \frac{\phi_B\left({\rm Tr}_A\left[ \rho_{AB}\, \phi_
		A^\star (E)\otimes \mathbbm{1}_B\right]\right)}{\tr{\rho_{AB} \, \phi_A^\star(E)\otimes \mathbbm{1}_B}}\right\} \nn \\
		&\subseteq\left\{ \frac{\phi_B\left({\rm Tr}_A\left[ \rho_{AB}\, E\otimes \mathbbm{1}_B\right]\right)}{\tr{\rho_{AB}\, E\otimes \mathbbm{1}_B}}\right\} \nn\\
		&=\left\{\phi_B\left( \frac{{\rm Tr}_A\left[ \rho_{AB}\,E\otimes \mathbbm{1}_B\right]}{\tr{\rho_{AB} \,E\otimes \mathbbm{1}_B}}\right)\right\} .\nn 		
	\end{align}
Hence, the steering ellipsoids of $\rho^\prime_{AB}$ and $\rho_{AB}$ are related by
\beq \label{subset}
{\cal E}'_{B|A}\subseteq \phi_B(\mathcal{E}_{B|A}). 
\eeq
	
To determine how the volumes of the steering ellipsoids are related, note that the trace distance between two states contracts under any CPTP map $\phi$~\cite{NC00}.
Moreover, for qubits, the trace distance is proportional to the Euclidean distance in the Bloch ball~\cite{NC00}.  As a consequence, the volume of any set of qubit Bloch vectors contracts under CPTP maps, yielding the inequality chain
	\beq
	V(\mathcal{E}^\prime_{B|A}) \leq V(\phi_B(\mathcal{E}_{B|A})) \leq V(\mathcal{E}_{B|A})
	\eeq
via Eq.~(\ref{subset}).
Thus,  local noise never increases the volume of the steering ellipsoid. 

 An immediate consequence of this result is that any volume monogamy relation, for a given set of multiqubit states, will remain valid under the addition of local noise. For example, it follows via Eq.~(\ref{monogamy}) that
	\beq
	\rt{v^\prime_{B|A}}+\rt{v^\prime_{C|A}} \leq 1 
		\eeq 
for any state obtained by adding local noise to a pure 3-qubit state, i.e., for any 3-qubit state of the form $\rho^\prime_{ABC}=(\phi_A\otimes\phi_B\otimes\phi_C)(|\psi_{ABC}\rangle\langle\psi_{ABC}|)$.

	\subsection{An example: local isotropic noise}

 We consider the set of states 
	\beq \label{example}
	\ket{\varphi_{ABC}}=p\ket{100}+\rt{\frac{1-p^2}{2}}\ket{010}+\rt{\frac{1-p^2}{2}}\ket{001},
	\eeq
	with $p \in (0, 1)$.  These states are symmetric with respect to Bob and Charlie, so that $v_{B|A}=v_{C|A}$. Moreover, it is easy to verify that these are W-class states~\cite{DVC00}.  Thus, they saturate the volume monogamy relation in Eq.~(\ref{monogamy}) (see Sec.~III~B), making them of experimental interest. 
	
However, the inevitable presence of noise in any experiment means that in practice these states cannot be perfectly generated and measured. We therefore investigate the robustness of these states  under a simple noise model. In particular, we consider the \blk  addition of local isotropic noise,
\beq
\rho^\prime_{ABC}=\Phi_\varepsilon(\rho_{ABC}):=(\phi_\varepsilon \otimes \phi_\varepsilon \otimes \phi_\varepsilon) (\rho_{ABC}), \label{noisy}
\eeq  
where
	\beq
	\phi_\varepsilon: \rho ~\rightarrow~~\frac{\varepsilon}{2}\mathbbm{1}+(1-\varepsilon)\rho 
  \eeq
 corresponds to adding isotropic noise of strength $\varepsilon \in [0, 1]$. Thus, for each qubit, $\varepsilon=0$ corresponds to no noise, while $\varepsilon=1$ corresponds to noise so strong that the state becomes completely mixed. The Bloch vector of each qubit is scaled by $1-\varepsilon$. 
  
\begin{figure}[t] \label{Fig2}
	\centering
	\includegraphics[width=0.45\textwidth]{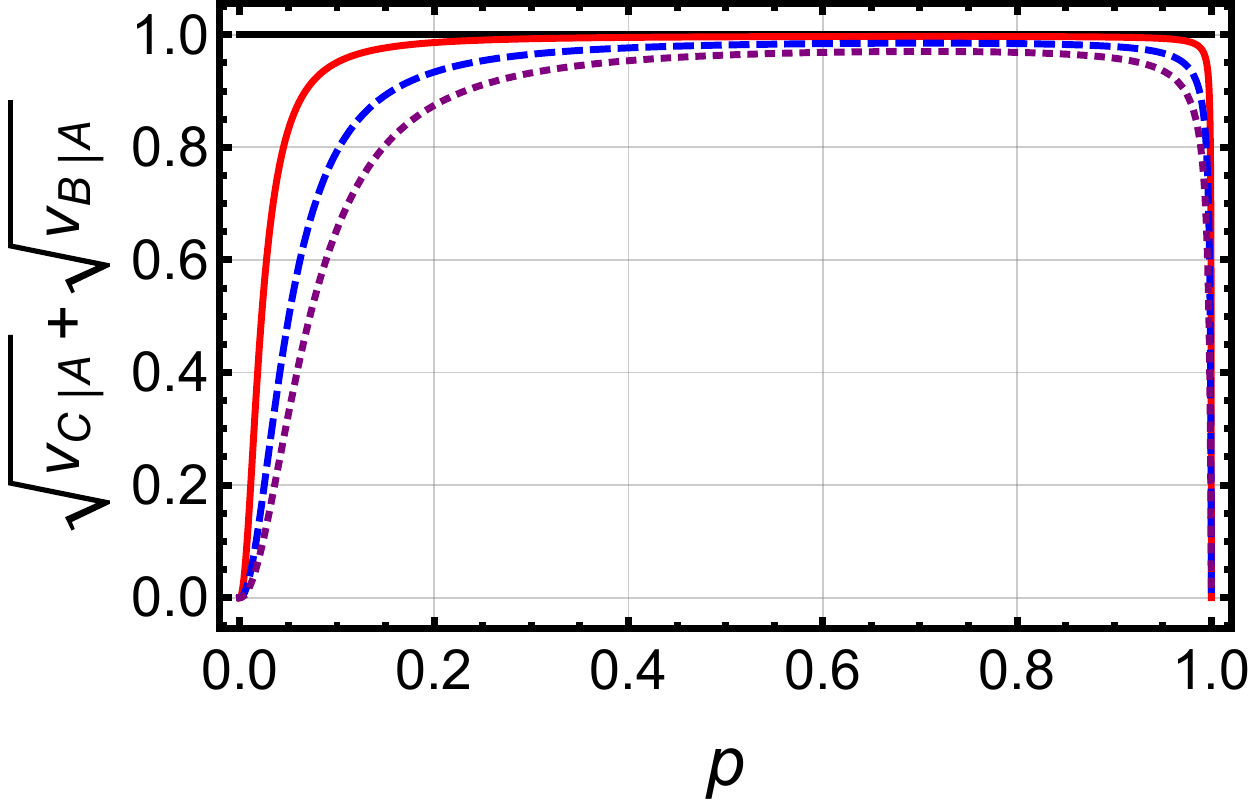}	
	\caption{ The effect of local noise on the left-hand-side of the volume monogamy relation, \erf{monogamy}, 
		for the family of states in \erf{example}.  
		Here, the noise on each qubit is a depolarizing channel with  strength \blk $\varepsilon=0~({\rm black~ solid~curve}),$ $ 0.001~({\rm red~solid~curve}),$ $ 0.005~({\rm blue~dashed~curve}),$ $0.01~({\rm purple~dotted~curve})$. }
\end{figure}

The noisy channel $\Phi_\varepsilon$ preserves the symmetry of the state $\rho_{ABC}$ with respect to Bob and Charlie, and the volumes of the steering ellipsoid may be analytically calculated for the states in Eq.~(\ref{example}) as
\beq
{v}^\prime_{B|A}=v^\prime_{C|A}=\frac{4p^4(1-p^2)^2(1-\varepsilon)^6}{[1-(1-\varepsilon)^2(1-2p^2)^2]^2}.
\eeq
\\ The corresponding sensitivity of the volume monogamy relation (\ref{monogamy}) to noise is depicted in Fig.~\ref{Fig2}, for a range of experimentally relevant noise strengths. It is seen that while the relation is no longer saturated for $\varepsilon>0$, those states with $p$ taking values in the midrange of the unit interval are relatively robust.

	\section{Conclusion} \label{V}
 We have studied volume monogamy relations for multi-qubit  systems. We have demonstrated a close connection between the volume monogamy relation~(\ref{monogamy}) for pure 3-qubit states and the SLOCC classification of such states.  A counterexample (\ref{counterexample}) was constructed to show this relation does not generalise to all 3-qubit states, and a suitable universal volume monogamy relation (\ref{analytic}) was obtained for the general case.  A similar relation was obtained in Eq.~(\ref{4qubit}) for  pure 4-qubit states, and conjectured to also hold for mixed states.\blk   ~Furthermore, we have found a generalised volume monogamy relation valid for all \blk multi-qubit states. 
 Finally, we studied the effects of noise on the quantum steering ellipsoid and showed that local noise channels do not invalidate the volume monogamy relation, as such noise decreases the volume of steering ellipsoids.  
  
 More generally, it is remarkable that the simple concept of the steering ellipsoid, i.e., the set of Bob's local states that Alice can prepare by local measurements on her system, can geometrically capture many important aspects of quantum correlations and information processing tasks.  Properties of steering ellipsoids are not only strongly connected to quantum monogamy, as investigated here, but have also been closely linked with quantum communication protocols based on, for example, Bell nonlocality \cite{MJJR14}, teleportation \cite{MJJR14}, and  Einstein-Podolsky-Rosen steering~\cite{SMMMH15, NV16S}.
  
 It is hoped that further investigation of steering ellipsoids (and their generalisations to higher dimensions) will cement their relevance to understanding the properties and usefulness of quantum correlations.  This goes well beyond properties of ellipsoid volumes: for example, while local dissipation reduces volumes as per Sec.~IV, it is also known that such noise can enhance teleportation fidelity for some states \cite{ BHHH00}.  Hence, a geometric characterisation of such states will require consideration of aspects other than volume (such as semiaxis lengths \cite{MJJR14}).
  
 Many open questions remain for future work even within the confines of volume monogamy relations. For example, \blk can stronger monogamy relations than Eqs.~(\ref{analytic}) and (\ref{4qubit}) be obtained?  Is  the 4-qubit conjecture in Eq.~(\ref{conjecture}) valid? \blk Is there some underlying connection between volume monogamy relations and that of other types of entanglement ~\cite{RDLA14, BXW14, ES15}?  How close might we get to these bounds with experiments? Finally, can volume monogamy be generalised to higher-dimensional systems?

	\acknowledgments
 We thank Jianxin Chen for his assistance in finding the counterexample~ (\ref{counterexample}). S. C. thanks Li Li for helpful discussions. This work was supported by the ARC Centre of Excellence CE110001027. A. M. is supported by the EPSRC.

	\bibliographystyle{apsrev4-1}
	\bibliography{monogamy}

\begin{thebibliography}{28}%
\makeatletter
\providecommand \@ifxundefined [1]{%
 \@ifx{#1\undefined}
}%
\providecommand \@ifnum [1]{%
 \ifnum #1\expandafter \@firstoftwo
 \else \expandafter \@secondoftwo
 \fi
}%
\providecommand \@ifx [1]{%
 \ifx #1\expandafter \@firstoftwo
 \else \expandafter \@secondoftwo
 \fi
}%
\providecommand \natexlab [1]{#1}%
\providecommand \enquote  [1]{``#1''}%
\providecommand \bibnamefont  [1]{#1}%
\providecommand \bibfnamefont [1]{#1}%
\providecommand \citenamefont [1]{#1}%
\providecommand \href@noop [0]{\@secondoftwo}%
\providecommand \href [0]{\begingroup \@sanitize@url \@href}%
\providecommand \@href[1]{\@@startlink{#1}\@@href}%
\providecommand \@@href[1]{\endgroup#1\@@endlink}%
\providecommand \@sanitize@url [0]{\catcode `\\12\catcode `\$12\catcode
  `\&12\catcode `\#12\catcode `\^12\catcode `\_12\catcode `\%12\relax}%
\providecommand \@@startlink[1]{}%
\providecommand \@@endlink[0]{}%
\providecommand \url  [0]{\begingroup\@sanitize@url \@url }%
\providecommand \@url [1]{\endgroup\@href {#1}{\urlprefix }}%
\providecommand \urlprefix  [0]{URL }%
\providecommand \Eprint [0]{\href }%
\providecommand \doibase [0]{http://dx.doi.org/}%
\providecommand \selectlanguage [0]{\@gobble}%
\providecommand \bibinfo  [0]{\@secondoftwo}%
\providecommand \bibfield  [0]{\@secondoftwo}%
\providecommand \translation [1]{[#1]}%
\providecommand \BibitemOpen [0]{}%
\providecommand \bibitemStop [0]{}%
\providecommand \bibitemNoStop [0]{.\EOS\space}%
\providecommand \EOS [0]{\spacefactor3000\relax}%
\providecommand \BibitemShut  [1]{\csname bibitem#1\endcsname}%
\let\auto@bib@innerbib\@empty
\bibitem [{\citenamefont {Nielsen}\ and\ \citenamefont {Chuang}(2010)}]{NC00}%
  \BibitemOpen
  \bibfield  {author} {\bibinfo {author} {\bibfnamefont {M.~A.}\ \bibnamefont
  {Nielsen}}\ and\ \bibinfo {author} {\bibfnamefont {I.~L.}\ \bibnamefont
  {Chuang}},\ }\href@noop {} {\emph {\bibinfo {title} {Quantum Computation and
  Quantum Information}}}\ (\bibinfo  {publisher} {Cambridge University Press},\
  \bibinfo {year} {2010})\BibitemShut {NoStop}%
\bibitem [{\citenamefont {Wiseman}\ and\ \citenamefont {Milburn}(2009)}]{WM09}%
  \BibitemOpen
  \bibfield  {author} {\bibinfo {author} {\bibfnamefont {H.~M.}\ \bibnamefont
  {Wiseman}}\ and\ \bibinfo {author} {\bibfnamefont {G.~J.}\ \bibnamefont
  {Milburn}},\ }\href@noop {} {\emph {\bibinfo {title} {Quantum Measurement and
  Control}}}\ (\bibinfo  {publisher} {Cambridge University Press},\ \bibinfo
  {year} {2009})\BibitemShut {NoStop}%
\bibitem [{\citenamefont {Jevtic}\ \emph {et~al.}(2014)\citenamefont {Jevtic},
  \citenamefont {Pusey}, \citenamefont {Jennings},\ and\ \citenamefont
  {Rudolph}}]{JPJR14}%
  \BibitemOpen
  \bibfield  {author} {\bibinfo {author} {\bibfnamefont {S.}~\bibnamefont
  {Jevtic}}, \bibinfo {author} {\bibfnamefont {M.}~\bibnamefont {Pusey}},
  \bibinfo {author} {\bibfnamefont {D.}~\bibnamefont {Jennings}}, \ and\
  \bibinfo {author} {\bibfnamefont {T.}~\bibnamefont {Rudolph}},\ }\href
  {\doibase 10.1103/PhysRevLett.113.020402} {\bibfield  {journal} {\bibinfo
  {journal} {Phys. Rev. Lett.}\ }\textbf {\bibinfo {volume} {113}},\ \bibinfo
  {pages} {020402} (\bibinfo {year} {2014})}\BibitemShut {NoStop}%
\bibitem [{\citenamefont {Shi}\ \emph {et~al.}(2011)\citenamefont {Shi},
  \citenamefont {Jiang}, \citenamefont {Sun},\ and\ \citenamefont
  {Du}}]{SJSD11}%
  \BibitemOpen
  \bibfield  {author} {\bibinfo {author} {\bibfnamefont {M.}~\bibnamefont
  {Shi}}, \bibinfo {author} {\bibfnamefont {F.}~\bibnamefont {Jiang}}, \bibinfo
  {author} {\bibfnamefont {C.}~\bibnamefont {Sun}}, \ and\ \bibinfo {author}
  {\bibfnamefont {J.}~\bibnamefont {Du}},\ }\href
  {http://stacks.iop.org/1367-2630/13/i=7/a=073016} {\bibfield  {journal}
  {\bibinfo  {journal} {New J. Phys.}\ }\textbf {\bibinfo {volume} {13}},\
  \bibinfo {pages} {073016} (\bibinfo {year} {2011})}\BibitemShut {NoStop}%
\bibitem [{\citenamefont {Shi}\ \emph {et~al.}(2012)\citenamefont {Shi},
  \citenamefont {Sun}, \citenamefont {Jiang}, \citenamefont {Yan},\ and\
  \citenamefont {Du}}]{SSJYD12}%
  \BibitemOpen
  \bibfield  {author} {\bibinfo {author} {\bibfnamefont {M.}~\bibnamefont
  {Shi}}, \bibinfo {author} {\bibfnamefont {C.}~\bibnamefont {Sun}}, \bibinfo
  {author} {\bibfnamefont {F.}~\bibnamefont {Jiang}}, \bibinfo {author}
  {\bibfnamefont {X.}~\bibnamefont {Yan}}, \ and\ \bibinfo {author}
  {\bibfnamefont {J.}~\bibnamefont {Du}},\ }\href {\doibase
  10.1103/PhysRevA.85.064104} {\bibfield  {journal} {\bibinfo  {journal} {Phys.
  Rev. A}\ }\textbf {\bibinfo {volume} {85}},\ \bibinfo {pages} {064104}
  (\bibinfo {year} {2012})}\BibitemShut {NoStop}%
\bibitem [{\citenamefont {Milne}\ \emph
  {et~al.}(2014{\natexlab{a}})\citenamefont {Milne}, \citenamefont {Jevtic},
  \citenamefont {Jennings}, \citenamefont {Wiseman},\ and\ \citenamefont
  {Rudolph}}]{MJJWR14}%
  \BibitemOpen
  \bibfield  {author} {\bibinfo {author} {\bibfnamefont {A.}~\bibnamefont
  {Milne}}, \bibinfo {author} {\bibfnamefont {S.}~\bibnamefont {Jevtic}},
  \bibinfo {author} {\bibfnamefont {D.}~\bibnamefont {Jennings}}, \bibinfo
  {author} {\bibfnamefont {H.}~\bibnamefont {Wiseman}}, \ and\ \bibinfo
  {author} {\bibfnamefont {T.}~\bibnamefont {Rudolph}},\ }\href
  {http://stacks.iop.org/1367-2630/16/i=8/a=083017} {\bibfield  {journal}
  {\bibinfo  {journal} {New J. Phys.}\ }\textbf {\bibinfo {volume} {16}},\
  \bibinfo {pages} {083017} (\bibinfo {year} {2014}{\natexlab{a}})}\BibitemShut
  {NoStop}%
\bibitem [{\citenamefont {Milne}\ \emph
  {et~al.}(2014{\natexlab{b}})\citenamefont {Milne}, \citenamefont {Jennings},
  \citenamefont {Jevtic},\ and\ \citenamefont {Rudolph}}]{MJJR14}%
  \BibitemOpen
  \bibfield  {author} {\bibinfo {author} {\bibfnamefont {A.}~\bibnamefont
  {Milne}}, \bibinfo {author} {\bibfnamefont {D.}~\bibnamefont {Jennings}},
  \bibinfo {author} {\bibfnamefont {S.}~\bibnamefont {Jevtic}}, \ and\ \bibinfo
  {author} {\bibfnamefont {T.}~\bibnamefont {Rudolph}},\ }\href {\doibase
  10.1103/PhysRevA.90.024302} {\bibfield  {journal} {\bibinfo  {journal} {Phys.
  Rev. A}\ }\textbf {\bibinfo {volume} {90}},\ \bibinfo {pages} {024302}
  (\bibinfo {year} {2014}{\natexlab{b}})}\BibitemShut {NoStop}%
\bibitem [{\citenamefont {Jevtic}\ \emph {et~al.}(2015)\citenamefont {Jevtic},
  \citenamefont {Hall}, \citenamefont {Anderson}, \citenamefont {Zwierz},\ and\
  \citenamefont {Wiseman}}]{SMMMH15}%
  \BibitemOpen
  \bibfield  {author} {\bibinfo {author} {\bibfnamefont {S.}~\bibnamefont
  {Jevtic}}, \bibinfo {author} {\bibfnamefont {M.~J.~W.}\ \bibnamefont {Hall}},
  \bibinfo {author} {\bibfnamefont {M.~R.}\ \bibnamefont {Anderson}}, \bibinfo
  {author} {\bibfnamefont {M.}~\bibnamefont {Zwierz}}, \ and\ \bibinfo {author}
  {\bibfnamefont {H.~M.}\ \bibnamefont {Wiseman}},\ }\href {\doibase
  10.1364/JOSAB.32.000A40} {\bibfield  {journal} {\bibinfo  {journal} {J. Opt.
  Soc. Am. B}\ }\textbf {\bibinfo {volume} {32}},\ \bibinfo {pages} {A40}
  (\bibinfo {year} {2015})}\BibitemShut {NoStop}%
\bibitem [{\citenamefont {Quan}\ \emph {et~al.}(2016)\citenamefont {Quan},
  \citenamefont {Zhu}, \citenamefont {Liu}, \citenamefont {Fei}, \citenamefont
  {Fan},\ and\ \citenamefont {Yang}}]{Quan16}%
  \BibitemOpen
  \bibfield  {author} {\bibinfo {author} {\bibfnamefont {Q.}~\bibnamefont
  {Quan}}, \bibinfo {author} {\bibfnamefont {H.}~\bibnamefont {Zhu}}, \bibinfo
  {author} {\bibfnamefont {S.-Y.}\ \bibnamefont {Liu}}, \bibinfo {author}
  {\bibfnamefont {S.-M.}\ \bibnamefont {Fei}}, \bibinfo {author} {\bibfnamefont
  {H.}~\bibnamefont {Fan}}, \ and\ \bibinfo {author} {\bibfnamefont {W.-L.}\
  \bibnamefont {Yang}},\ }\href {http://dx.doi.org/10.1038/srep22025}
  {\bibfield  {journal} {\bibinfo  {journal} {Sci. Rep.}\ }\textbf {\bibinfo
  {volume} {6}},\ \bibinfo {pages} {22025} (\bibinfo {year}
  {2016})}\BibitemShut {NoStop}%
\bibitem [{\citenamefont {Nguyen}\ and\ \citenamefont
  {Vu}(2016{\natexlab{a}})}]{NV16F}%
  \BibitemOpen
  \bibfield  {author} {\bibinfo {author} {\bibfnamefont {H.~C.}\ \bibnamefont
  {Nguyen}}\ and\ \bibinfo {author} {\bibfnamefont {T.}~\bibnamefont {Vu}},\
  }\href {\doibase 10.1103/PhysRevA.94.012114} {\bibfield  {journal} {\bibinfo
  {journal} {Phys. Rev. A}\ }\textbf {\bibinfo {volume} {94}},\ \bibinfo
  {pages} {012114} (\bibinfo {year} {2016}{\natexlab{a}})}\BibitemShut
  {NoStop}%
\bibitem [{\citenamefont {Nguyen}\ and\ \citenamefont
  {Vu}(2016{\natexlab{b}})}]{NV16S}%
  \BibitemOpen
  \bibfield  {author} {\bibinfo {author} {\bibfnamefont {H.~C.}\ \bibnamefont
  {Nguyen}}\ and\ \bibinfo {author} {\bibfnamefont {T.}~\bibnamefont {Vu}},\
  }\href {http://stacks.iop.org/0295-5075/115/i=1/a=10003} {\bibfield
  {journal} {\bibinfo  {journal} {Europhys. Lett.}\ }\textbf {\bibinfo {volume}
  {115}},\ \bibinfo {pages} {10003} (\bibinfo {year}
  {2016}{\natexlab{b}})}\BibitemShut {NoStop}%
\bibitem [{\citenamefont {Coffman}\ \emph {et~al.}(2000)\citenamefont
  {Coffman}, \citenamefont {Kundu},\ and\ \citenamefont {Wootters}}]{CKW00}%
  \BibitemOpen
  \bibfield  {author} {\bibinfo {author} {\bibfnamefont {V.}~\bibnamefont
  {Coffman}}, \bibinfo {author} {\bibfnamefont {J.}~\bibnamefont {Kundu}}, \
  and\ \bibinfo {author} {\bibfnamefont {W.~K.}\ \bibnamefont {Wootters}},\
  }\href {\doibase 10.1103/PhysRevA.61.052306} {\bibfield  {journal} {\bibinfo
  {journal} {Phys. Rev. A}\ }\textbf {\bibinfo {volume} {61}},\ \bibinfo
  {pages} {052306} (\bibinfo {year} {2000})}\BibitemShut {NoStop}%
\bibitem [{\citenamefont {Osborne}\ and\ \citenamefont
  {Verstraete}(2006)}]{OV06}%
  \BibitemOpen
  \bibfield  {author} {\bibinfo {author} {\bibfnamefont {T.~J.}\ \bibnamefont
  {Osborne}}\ and\ \bibinfo {author} {\bibfnamefont {F.}~\bibnamefont
  {Verstraete}},\ }\href {\doibase 10.1103/PhysRevLett.96.220503} {\bibfield
  {journal} {\bibinfo  {journal} {Phys. Rev. Lett.}\ }\textbf {\bibinfo
  {volume} {96}},\ \bibinfo {pages} {220503} (\bibinfo {year}
  {2006})}\BibitemShut {NoStop}%
\bibitem [{\citenamefont {Scarani}\ and\ \citenamefont {Gisin}(2001)}]{SG01}%
  \BibitemOpen
  \bibfield  {author} {\bibinfo {author} {\bibfnamefont {V.}~\bibnamefont
  {Scarani}}\ and\ \bibinfo {author} {\bibfnamefont {N.}~\bibnamefont
  {Gisin}},\ }\href {\doibase 10.1103/PhysRevLett.87.117901} {\bibfield
  {journal} {\bibinfo  {journal} {Phys. Rev. Lett.}\ }\textbf {\bibinfo
  {volume} {87}},\ \bibinfo {pages} {117901} (\bibinfo {year}
  {2001})}\BibitemShut {NoStop}%
\bibitem [{\citenamefont {Toner}\ and\ \citenamefont
  {Verstraete}(2006)}]{TV06}%
  \BibitemOpen
  \bibfield  {author} {\bibinfo {author} {\bibfnamefont {B.}~\bibnamefont
  {Toner}}\ and\ \bibinfo {author} {\bibfnamefont {F.}~\bibnamefont
  {Verstraete}},\ }\href@noop {} {\bibfield  {journal} {\bibinfo  {journal}
  {arXiv: quant-ph/0611001}\ } (\bibinfo {year} {2006})}\BibitemShut {NoStop}%
\bibitem [{\citenamefont {Toner}(2009)}]{T09}%
  \BibitemOpen
  \bibfield  {author} {\bibinfo {author} {\bibfnamefont {B.}~\bibnamefont
  {Toner}},\ }\href {\doibase 10.1098/rspa.2008.0149} {\bibfield  {journal}
  {\bibinfo  {journal} {Proc. R. Soc. A}\ }\textbf {\bibinfo {volume} {465}},\
  \bibinfo {pages} {59} (\bibinfo {year} {2009})}\BibitemShut {NoStop}%
\bibitem [{\citenamefont {Kurzy\ifmmode~\acute{n}\else \'{n}\fi{}ski}\ \emph
  {et~al.}(2011)\citenamefont {Kurzy\ifmmode~\acute{n}\else \'{n}\fi{}ski},
  \citenamefont {Paterek}, \citenamefont {Ramanathan}, \citenamefont
  {Laskowski},\ and\ \citenamefont {Kaszlikowski}}]{KPRLK11}%
  \BibitemOpen
  \bibfield  {author} {\bibinfo {author} {\bibfnamefont {P.}~\bibnamefont
  {Kurzy\ifmmode~\acute{n}\else \'{n}\fi{}ski}}, \bibinfo {author}
  {\bibfnamefont {T.}~\bibnamefont {Paterek}}, \bibinfo {author} {\bibfnamefont
  {R.}~\bibnamefont {Ramanathan}}, \bibinfo {author} {\bibfnamefont
  {W.}~\bibnamefont {Laskowski}}, \ and\ \bibinfo {author} {\bibfnamefont
  {D.}~\bibnamefont {Kaszlikowski}},\ }\href {\doibase
  10.1103/PhysRevLett.106.180402} {\bibfield  {journal} {\bibinfo  {journal}
  {Phys. Rev. Lett.}\ }\textbf {\bibinfo {volume} {106}},\ \bibinfo {pages}
  {180402} (\bibinfo {year} {2011})}\BibitemShut {NoStop}%
\bibitem [{\citenamefont {Reid}(2013)}]{R13}%
  \BibitemOpen
  \bibfield  {author} {\bibinfo {author} {\bibfnamefont {M.~D.}\ \bibnamefont
  {Reid}},\ }\href {\doibase 10.1103/PhysRevA.88.062108} {\bibfield  {journal}
  {\bibinfo  {journal} {Phys. Rev. A}\ }\textbf {\bibinfo {volume} {88}},\
  \bibinfo {pages} {062108} (\bibinfo {year} {2013})}\BibitemShut {NoStop}%
\bibitem [{\citenamefont {Klyachko}(2004)}]{K04}%
  \BibitemOpen
  \bibfield  {author} {\bibinfo {author} {\bibfnamefont {A.}~\bibnamefont
  {Klyachko}},\ }\href@noop {} {\bibfield  {journal} {\bibinfo  {journal}
  {arXiv: quant-ph/0409113}\ } (\bibinfo {year} {2004})}\BibitemShut {NoStop}%
\bibitem [{\citenamefont {Hu}\ and\ \citenamefont {Fan}(2015)}]{HF15}%
  \BibitemOpen
  \bibfield  {author} {\bibinfo {author} {\bibfnamefont {X.}~\bibnamefont
  {Hu}}\ and\ \bibinfo {author} {\bibfnamefont {H.}~\bibnamefont {Fan}},\
  }\href {\doibase 10.1103/PhysRevA.91.022301} {\bibfield  {journal} {\bibinfo
  {journal} {Phys. Rev. A}\ }\textbf {\bibinfo {volume} {91}},\ \bibinfo
  {pages} {022301} (\bibinfo {year} {2015})}\BibitemShut {NoStop}%
\bibitem [{\citenamefont {Milne}\ \emph {et~al.}(2015)\citenamefont {Milne},
  \citenamefont {Jevtic}, \citenamefont {Jennings}, \citenamefont {Wiseman},\
  and\ \citenamefont {Rudolph}}]{ASDHT15}%
  \BibitemOpen
  \bibfield  {author} {\bibinfo {author} {\bibfnamefont {A.}~\bibnamefont
  {Milne}}, \bibinfo {author} {\bibfnamefont {S.}~\bibnamefont {Jevtic}},
  \bibinfo {author} {\bibfnamefont {D.}~\bibnamefont {Jennings}}, \bibinfo
  {author} {\bibfnamefont {H.}~\bibnamefont {Wiseman}}, \ and\ \bibinfo
  {author} {\bibfnamefont {T.}~\bibnamefont {Rudolph}},\ }\href
  {http://stacks.iop.org/1367-2630/17/i=1/a=019501} {\bibfield  {journal}
  {\bibinfo  {journal} {New J. Phys.}\ }\textbf {\bibinfo {volume} {17}},\
  \bibinfo {pages} {019501} (\bibinfo {year} {2015})}\BibitemShut {NoStop}%
\bibitem [{\citenamefont {Higuchi}\ \emph {et~al.}(2003)\citenamefont
  {Higuchi}, \citenamefont {Sudbery},\ and\ \citenamefont {Szulc}}]{HSS03}%
  \BibitemOpen
  \bibfield  {author} {\bibinfo {author} {\bibfnamefont {A.}~\bibnamefont
  {Higuchi}}, \bibinfo {author} {\bibfnamefont {A.}~\bibnamefont {Sudbery}}, \
  and\ \bibinfo {author} {\bibfnamefont {J.}~\bibnamefont {Szulc}},\ }\href
  {\doibase 10.1103/PhysRevLett.90.107902} {\bibfield  {journal} {\bibinfo
  {journal} {Phys. Rev. Lett.}\ }\textbf {\bibinfo {volume} {90}},\ \bibinfo
  {pages} {107902} (\bibinfo {year} {2003})}\BibitemShut {NoStop}%
\bibitem [{\citenamefont {D\"ur}\ \emph {et~al.}(2000)\citenamefont {D\"ur},
  \citenamefont {Vidal},\ and\ \citenamefont {Cirac}}]{DVC00}%
  \BibitemOpen
  \bibfield  {author} {\bibinfo {author} {\bibfnamefont {W.}~\bibnamefont
  {D\"ur}}, \bibinfo {author} {\bibfnamefont {G.}~\bibnamefont {Vidal}}, \ and\
  \bibinfo {author} {\bibfnamefont {J.~I.}\ \bibnamefont {Cirac}},\ }\href
  {\doibase 10.1103/PhysRevA.62.062314} {\bibfield  {journal} {\bibinfo
  {journal} {Phys. Rev. A}\ }\textbf {\bibinfo {volume} {62}},\ \bibinfo
  {pages} {062314} (\bibinfo {year} {2000})}\BibitemShut {NoStop}%
\bibitem [{\citenamefont {Qin}\ \emph {et~al.}(2015)\citenamefont {Qin},
  \citenamefont {Fei},\ and\ \citenamefont {Li-Jost}}]{QFL15}%
  \BibitemOpen
  \bibfield  {author} {\bibinfo {author} {\bibfnamefont {H.-H.}\ \bibnamefont
  {Qin}}, \bibinfo {author} {\bibfnamefont {S.-M.}\ \bibnamefont {Fei}}, \ and\
  \bibinfo {author} {\bibfnamefont {X.}~\bibnamefont {Li-Jost}},\ }\href
  {\doibase 10.1103/PhysRevA.92.062339} {\bibfield  {journal} {\bibinfo
  {journal} {Phys. Rev. A}\ }\textbf {\bibinfo {volume} {92}},\ \bibinfo
  {pages} {062339} (\bibinfo {year} {2015})}\BibitemShut {NoStop}%
\bibitem [{\citenamefont {Badzia\ifmmode~\mbox{\c{}}\else \c{}\fi{}g}\ \emph
  {et~al.}(2000)\citenamefont {Badzia\ifmmode~\mbox{\c{}}\else \c{}\fi{}g},
  \citenamefont {Horodecki}, \citenamefont {Horodecki},\ and\ \citenamefont
  {Horodecki}}]{BHHH00}%
  \BibitemOpen
  \bibfield  {author} {\bibinfo {author} {\bibfnamefont {P.}~\bibnamefont
  {Badzia\ifmmode~\mbox{\c{}}\else \c{}\fi{}g}}, \bibinfo {author}
  {\bibfnamefont {M.}~\bibnamefont {Horodecki}}, \bibinfo {author}
  {\bibfnamefont {P.}~\bibnamefont {Horodecki}}, \ and\ \bibinfo {author}
  {\bibfnamefont {R.}~\bibnamefont {Horodecki}},\ }\href {\doibase
  10.1103/PhysRevA.62.012311} {\bibfield  {journal} {\bibinfo  {journal} {Phys.
  Rev. A}\ }\textbf {\bibinfo {volume} {62}},\ \bibinfo {pages} {012311}
  (\bibinfo {year} {2000})}\BibitemShut {NoStop}%
\bibitem [{\citenamefont {Regula}\ \emph {et~al.}(2014)\citenamefont {Regula},
  \citenamefont {Di~Martino}, \citenamefont {Lee},\ and\ \citenamefont
  {Adesso}}]{RDLA14}%
  \BibitemOpen
  \bibfield  {author} {\bibinfo {author} {\bibfnamefont {B.}~\bibnamefont
  {Regula}}, \bibinfo {author} {\bibfnamefont {S.}~\bibnamefont {Di~Martino}},
  \bibinfo {author} {\bibfnamefont {S.}~\bibnamefont {Lee}}, \ and\ \bibinfo
  {author} {\bibfnamefont {G.}~\bibnamefont {Adesso}},\ }\href {\doibase
  10.1103/PhysRevLett.113.110501} {\bibfield  {journal} {\bibinfo  {journal}
  {Phys. Rev. Lett.}\ }\textbf {\bibinfo {volume} {113}},\ \bibinfo {pages}
  {110501} (\bibinfo {year} {2014})}\BibitemShut {NoStop}%
\bibitem [{\citenamefont {Bai}\ \emph {et~al.}(2014)\citenamefont {Bai},
  \citenamefont {Xu},\ and\ \citenamefont {Wang}}]{BXW14}%
  \BibitemOpen
  \bibfield  {author} {\bibinfo {author} {\bibfnamefont {Y.-K.}\ \bibnamefont
  {Bai}}, \bibinfo {author} {\bibfnamefont {Y.-F.}\ \bibnamefont {Xu}}, \ and\
  \bibinfo {author} {\bibfnamefont {Z.~D.}\ \bibnamefont {Wang}},\ }\href
  {\doibase 10.1103/PhysRevLett.113.100503} {\bibfield  {journal} {\bibinfo
  {journal} {Phys. Rev. Lett.}\ }\textbf {\bibinfo {volume} {113}},\ \bibinfo
  {pages} {100503} (\bibinfo {year} {2014})}\BibitemShut {NoStop}%
\bibitem [{\citenamefont {Eltschka}\ and\ \citenamefont
  {Siewert}(2015)}]{ES15}%
  \BibitemOpen
  \bibfield  {author} {\bibinfo {author} {\bibfnamefont {C.}~\bibnamefont
  {Eltschka}}\ and\ \bibinfo {author} {\bibfnamefont {J.}~\bibnamefont
  {Siewert}},\ }\href {\doibase 10.1103/PhysRevLett.114.140402} {\bibfield
  {journal} {\bibinfo  {journal} {Phys. Rev. Lett.}\ }\textbf {\bibinfo
  {volume} {114}},\ \bibinfo {pages} {140402} (\bibinfo {year}
  {2015})}\BibitemShut {NoStop}%
\end{thebibliography}%

\end{document}